\documentclass[aps,pre,superscriptaddress,onecolumn]{revtex4}
\usepackage{graphicx,amsmath,amssymb}

\newcommand{\aN}{b_N}
\newcommand{\bN}{a_N}
\newcommand{\alphaN}{\beta_N}

\begin{document}

\title{Accurately approximating extreme value statistics}

\author{Lior Zarfaty}
\affiliation{Department of Physics, Institute of Nanotechnology and Advanced Materials, Bar-Ilan University, Ramat-Gan 52900, Israel}

\author{Eli Barkai}
\affiliation{Department of Physics, Institute of Nanotechnology and Advanced Materials, Bar-Ilan University, Ramat-Gan 52900, Israel}

\author{David A. Kessler}
\affiliation{Department of Physics, Bar-Ilan University, Ramat-Gan 52900, Israel}

\begin{abstract}

We consider the extreme value statistics of $N$ independent and identically distributed random variables, which is a classic problem in probability theory. When $N\to\infty$, fluctuations around the maximum of the variables are described by the Fisher-Tippett-Gnedenko theorem, which states that the distribution of maxima converges to one out of three limiting forms. Among these is the Gumbel distribution, for which the convergence rate with $N$ is of a logarithmic nature. Here, we present a theory that allows one to use the Gumbel limit to accurately approximate the exact extreme value distribution. We do so by representing the scale and width parameters as power series, and by a transformation of the underlying distribution. We consider functional corrections to the Gumbel limit as well, showing they are obtainable via Taylor expansion. Our method also improves the description of large deviations from the mean extreme value. Additionally, it helps to characterize the extreme value statistics when the underlying distribution is unknown, for example when fitting experimental data.

\end{abstract}

\maketitle

\section{Introduction}

Extreme value (EV) statistics \cite{Gumbel,Leadbetter,Majumdar,Hansen} is an important subfield of probability theory. Given a random variable $\chi$ which describes the magnitude of a recurring event, the focus is on the statistical properties of the maximal value of a set of $N$ such events. Ever since the foundational work on this problem by Fisher and Tippett \cite{Fisher}, it has continued to attract interest. Problems involving EVs of a large number of random variables are important in many fields of physics \cite{Fortin}, such as brittle fracture \cite{Weibull,HuntMcCartney,Alava}, disordered systems \cite{BouchardMezard,Burioni,Barkai1}, $1/f$ noise \cite{Fyodorov}, renewal processes \cite{Schehr}, long-ranged Ising systems \cite{Mukamel}, condensation \cite{Godreche,Barkai2}, and galaxy clusters \cite{Silk}, as well as a broad range of other applications including meteorology \cite{Abarbanel}, finance \cite{Embrechts,Novak,TracyWidom}, and the immune system \cite{George}.

To formulate the problem under discussion, let $\{\chi_1,...,\chi_N\}$ be a set of $N$ independent and identically distributed (IID) unbounded random variables $\chi_i \in (-\infty,\infty)$, with a common cumulative distribution function (CDF) $F(\chi)$, and a probability density function (PDF) $f(\chi)\equiv\text{d}F/\text{d}\chi$ that decreases faster than a power-law for large $\chi$. The maximal value of this set, denoted as $x \equiv \max(\{\chi_1,...,\chi_N\})$, has an exact CDF of $F_N(x)=F^N(x)$. Note that a fixed $x=x_0$ together with an everywhere differentiable $F(\chi)$ leads to a single possible outcome, $\lim_{N\to\infty}F_N(x_0)=0$. However, increasing $x$ brings $F_N(x)$ closer to unity, and thus a nontrivial limit emerges upon taking $N,x\to\infty$ simultaneously. This is attainable by suitably choosing two scaling sequences $\aN$ and $\bN$, while rescaling $x$ as $z \equiv (x-\aN)/\bN$, $z \sim O(1)$, leading to a convergence in distribution of $F_N(\aN+\bN z)$. Namely,
\begin{equation}
\label{equation: primary definition extreme value}
	G_N(z) \equiv F_N \left( \aN + \bN z \right) , \quad \lim_{N\to\infty} G_N(z) = G_{\infty}(z) \equiv \exp\left(-e^{-z}\right) ,
\end{equation}
where $G_{\infty}(z)$ is the Gumbel CDF \cite{Fisher}. Therefore, $\aN$ and $\bN$ represent the location and width, respectively, of the EV distribution. Note that we designate the CDF (PDF) of the maximal value $x$ by $F_N$ ($f_N$), whereas distributions of the scaled variable $z$ are designated by $G_N$ ($g_N$), respectively. Importantly, the choice of $\aN$ and $\bN$ is not unique. A second set of sequences $\aN'$ and $\bN'$ can serve as an appropriate candidate if the following conditions hold \cite{Gnedenko}, 
\begin{equation}
\label{equation: conditions on sequences}
    \lim_{N\to\infty}\frac{\aN-\aN'}{\bN} = 0 , \quad \lim_{N\to\infty}\frac{\bN'}{\bN} = 1 ,
\end{equation}
so that the locations on the scale of the width, and the widths themselves, are asymptotically identical.

Even though the limit of $N\to\infty$ is long understood, the convergence rate to the Gumbel form is logarithmic in nature for any $f(\chi)$ which is not purely exponential, including the familiar Gaussian \cite{Hall}. It turns out that this convergence rate is extremely sensitive to the choice of $\aN$ and $\bN$, as we show below. Even worse, trying to approximate these sequences for large $N$ results in corrections that involve iterated-logarithm terms, preventing the usage of convergence acceleration techniques such as Pad\'e approximants. Hence, a power-series representation of these sequences can greatly assist in generating an accurate Gumbel approximation to $G_N(z)$ for large, but not exponentially large, $N$s. As one decreases $N$, one finds that no simple Gumbel approximation is satisfactory for even the best choice of $\aN$ and $\bN$, as the distribution $G_N(z)$ increasingly diverges from the asymptotic Gumbel form. One possible workaround is calculating functional corrections to $G_{\infty}(z)$ that allow for accurate capture of the true distribution $G_N(z)$, as we shall demonstrate. However, we first introduce a different method, which we find more efficient: generating the Gumbel approximation for a transformed variable, and using this to construct an approximate distribution for the original variable. In any case, it is clear that to make practical uses of the limit law in the Gumbel case, one always needs to have good estimates of the location and width, namely $\aN$ and $\bN$.

In their body of research, Gy\"{o}rgyi, et al. \cite{Gyorgyi1,Gyorgyi2,Gyorgyi3} explored this problem of finite $N$ using a renormalization-group approach. They found the first-order correction of $G_N$ to the Gumbel distribution, and showed that it has a universal structure. By universality, it is meant that this correction has a functional shape that is independent of the underlying distribution $F$, and the $F$-dependence enters only via a numerical prefactor to the functional correction. They also obtained explicit expressions for given general asymptotic shapes of $F$, and showed that the first-order correction contributes to convergence in certain correlated systems (i.e. percolation and $1/f$ noise). However, the importance of an accurate estimation of $\aN$ and $\bN$ was not discussed in these works. In addition, they restricted themselves to the first correction, and indeed obtaining higher order terms using the renormalization-group is not an easy task.

Our exposition, then, is comprised of two main parts. The first part centers around an optimal use of the Gumbel distribution, $G_{\infty}(z)$, without a need for functional corrections. It is based primarily on an approximation of the sequences $\aN$ and $\bN$ via power-series expansions, given a general model of stretched or compressed exponential distribution $F$, which also includes the Gaussian. These power series rely only on the behavior of $F(\chi)$ at $\chi\to\infty$, and are expressed in terms of a single large parameter $\alphaN$ that encapsulates all the complicated iterated logarithmic $N$-dependencies by means of the Lambert W-function. In addition, we make a simple change of variables that brings the underlying distribution more to an exponential-like shape, drastically accelerating the convergence rate. This yields closed-form expressions for $\aN$ and $\bN$, working excellently down to $N=50$ (or $500$ for extreme examples) for the scenarios we examined. Our theory speeds up convergence dramatically when compared to the simple $\ln(N)$ scaling of the sequences typically used \cite{Majumdar,Gyorgyi2}.

The second part is a procedure for deriving functional corrections of any order to the Gumbel distribution, done by Taylor expanding the double logarithm of the underlying distribution $F$. This process yields $G_N$ as depending on numerical coefficients expressed via $F$, providing an arbitrary-order expansion around $G_{\infty}$, and here we explicitly state the second correction. In agreement with Gy\"{o}rgyi, et al., we find that the first correction to the Gumbel distribution has a universal functional shape, with the methods of part one providing a much faster convergence. This part provides us with the ability to approximate the moments of the EV distribution to arbitrary precision.

Note that the limit in Eq.~(\ref{equation: primary definition extreme value}) implicitly assumes that $z \sim O(1)$, and thus for finite $N$ the Gumbel form approximates only the bulk of the exact EV distribution $F_N(x)$. To accurately describe the right tail of $F_N(x)$, one needs to exploit large deviation theory \cite{Touchette}. Using a different pair of scaling sequences $s_N$ and $u_N$, one defines
\begin{equation}
\label{equation: primary definition large deviations 1}
	H_N(\xi) \equiv 1-F_N \left( s_N \xi \right) , \quad \psi(\xi) \equiv -\lim_{N\to\infty} \frac{1}{u_N}\ln\left[H_N(\xi)\right] ,
\end{equation}
so that
\begin{equation}
\label{equation: primary definition large deviations 2}
	H_N(\xi) \approx e^{-u_N\psi(\xi)} \text{ for } N \gg 1 \text{ and } \xi \ge 1
\end{equation}
at the distribution's right tail \cite{Rita}. Traditionally, $u_N$ is called the speed and $\psi(\xi)$ is termed the rate function \cite{Vivo}. Usually for large deviations, a $1/N$ scaling is used for the rescaled variable $\xi$, but here a different scaling needs to be applied, $\xi=x/s_N$, with $s_N$ to be determined \cite{Rita,Vivo}. However, the resulting theory suffers from the same convergence problem mentioned for the typical fluctuations of the maximum.

Hence, we consider the large deviations regime as well, where we find that reexpressing the $N$-dependence in terms of $\alphaN$ (using the Lambert W-function) resolves this domain's convergence problem. The left tail is more challenging and does not possess a simple large deviation form to the best of our knowledge. Nevertheless, we derive a uniform approximation describing it. It can be regarded as an extreme large deviations principle, in which the PDF's {\em double} logarithm has a large deviation form.

We also consider the EV problem from a practical data analysis direction, where we demonstrate that our approach does not require any knowledge of the underlying distribution. Given a data set of maxima which, in principle, is attracted to the Gumbel law in the limit of $N\to\infty$, we describe an algorithm that can be used to extract the EV distribution parameters ($\aN$, $\bN$, and the Taylor coefficients responsible for the functional corrections), while accounting for the change of variables method, and show it works for $N$s as small as $25$ for various examples of the underlying CDF.

Finally, we discuss other cases of EVs. Firstly, we deal with the problem of the fastest first-return time \cite{Schuss,Lawley}, which is a case of minimal EV statistics with a lower bound, that nevertheless has a Gumbel limit which is approached extremely slowly. Secondly, we briefly consider the other two EV limits, the Fr\'echet and Weibull distributions, showing how their underlying CDFs can be usefully understood as an exponential CDF of a transformed variable, thereby shedding light on the reason for which the convergence of underlying distributions to these limits is much faster.

The rest of this paper is organized as follows. In Sec.~\ref{section: uncorrected} we develop our theory that allows for utilization of the Gumbel limit, namely $G_{\infty}(z)$, to accurately predict the EV PDF. We obtain expansions to the sequences $\aN$ and $\bN$ given a general asymptotic behavior of $F$, working down to $N=50$. In Sec.~\ref{section: corrections} we outline our method for deriving arbitrary-order corrections to the Gumbel distribution, obtaining expressions for the first two corrections, and observing their shape. In Sec.~\ref{section: tails} we provide a treatment of the far tails. In Sec.~\ref{section: practical} we discuss the EV statistics from a practical data analysis point of view, presenting a fitting-based method that works when the underlying distribution $F$ is not known. Sec.~\ref{section: others} is dedicated to other cases of EV problems, the minimum case alluded to above and the other two EV limits, the Fr\'echet and Weibull distributions. Lastly, we summarize our results in Sec.~\ref{section: summary}.

\section{Fast convergence to the Gumbel limit}
\label{section: uncorrected}

As stated in the introduction, our primary aim is to obtain an accurate approximation to the EV distribution $f_N(x)$. We lay the foundations for our theory by assuming that the leading large-$\chi$ asymptotic behavior of the common CDF $F(\chi)$ is known. We employ a combination of two techniques for accurately approximating the aforementioned EV PDF.

The first one allows for an accurate evaluation of the scaling sequences $\aN$ and $\bN$. As shown below, these can in principle be determined via an inversion of the exact underlying CDF $F(\chi)$, which is assumed here to not be explicitly known. Moreover, even given $F(\chi)$, the $N$ dependence of these parameters is extremely complicated, precluding analytical progress. We present a method that accurately approximates the exact values of these sequences in terms of the Lambert W-function.

Nevertheless, for certain types of common distributions, this is not enough, as the convergence rate is inherently even slower than usual. These cases are characterized by being ``far" from an exponential distribution, a characterization on which we elaborate below in more detail (e.g. a very stretched exponential falls into this category). Here enters our second technique: by performing a transformation of variables aimed at making the underlying CDF more similar to the rapidly converging exponential case, we make the Gumbel limit usable when combined with the first method discussed above.

\subsection{Approximating $\aN$ and $\bN$}

We start our calculations following Gy\"{o}rgyi \cite{Gyorgyi2}, by rewriting $F(\chi)$ as
\begin{equation}
\label{equation: l definition}
	F(\chi) \equiv \exp\left\{-\exp\left[-L(\chi)\right]\right\} ,
\end{equation}
so that $F_N(x)=\exp\left\{-N\exp\left[-L(x)\right]\right\}$. The advantage of this representation is that in the center part of the EV distribution, $L$ can be replaced by a low-order polynomial, and the larger $N$ is, the smaller is the higher-order terms' impact. Plugging in $x=\aN+\bN z$ and assuming that $\aN\gg\bN$ with $z\sim O(1)$, such that $x\approx\aN$, we can expand
\begin{equation}
\label{equation: l expansion}
	L(x) = \sum_{n=0}^{\infty} \frac{1}{n!}L^{(n)}(\aN)(\bN z)^n = \sum_{n=0}^{\infty} \frac{c_n}{n!}z^n ,
\end{equation}
where $L^{(n)}(x)\equiv\text{d}^nL(x)/\text{d}x^n$. As we are interested with the Gumbel limit, it is natural to define the scaling sequences as
\begin{equation}
\label{equation: normalization sequences definition}
	\exp\left[L(\aN)\right] = N , \quad \bN = \frac{1}{L^{(1)}(\aN)} , \quad c_n(\aN) \equiv \frac{L^{(n)}(\aN)}{[L^{(1)}(\aN)]^n} ,
\end{equation}
since then $G_N(z) = \exp\{-\exp[-z+O(c_2)]\}$. The key point is that for the broad class of generalized (stretched or compressed) exponential distributions, $L(\chi) \propto \chi^\nu$ with $\nu>0$, so that $\aN \propto [\ln(N)]^{1/\nu} \gg 1$ and $c_2 \propto 1/\aN^{\nu} \ll 1$ for large $N$. While this particular choice of $\aN$ and $\bN$ has a degree of arbitrariness, as explained above, it is crucial that any approximation of $\aN$, which we can denote by $\aN'$, satisfies that $|\aN-\aN'|/\bN$ be reasonably small, say less than $0.1$, for all $N$s of interest. We shall now see that this is not true for the naive large-$N$ approximation defined below, henceforth referred to as the ``standard" approximation, even though it is true asymptotically for extremely large $N$. Our first task will be to address this challenge.
\begin{figure}
    \includegraphics[width=1.0\textwidth]{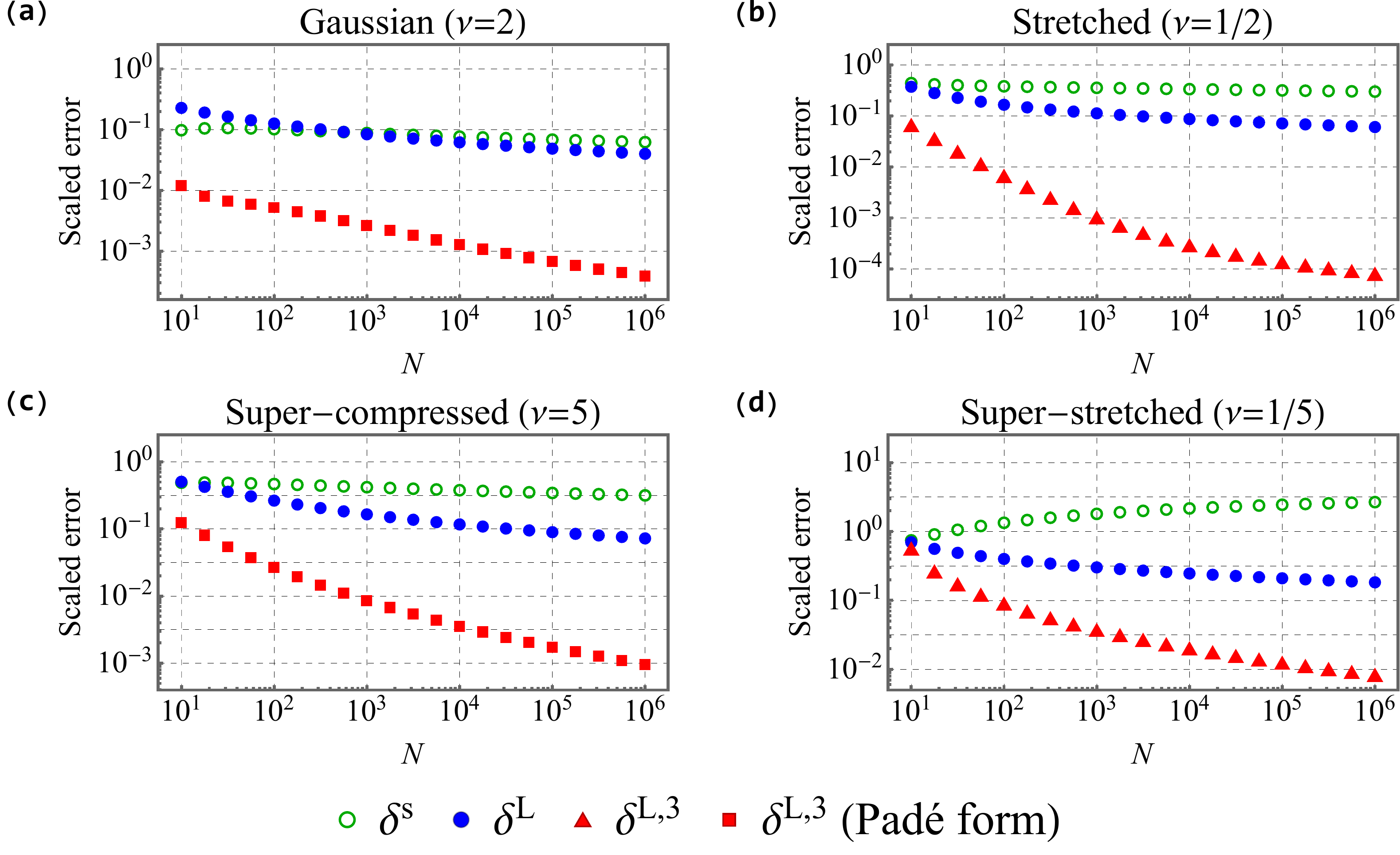}
    \caption{The scaled error in $\aN$, relative to the width $\bN$, $\delta \equiv |\aN-\aN'|/\bN$, where $\aN'$ is given for the standard approximation by Eq.~(\ref{equation: sequences solved simple}) (green circles), for the Lambert approximation by Eq.~(\ref{equation: normalization sequences zero order}) (blue disks), and for the series expansions by Eq.~(\ref{equation: normalization sequences expansion}) (red triangles, red squares when $[1/2]$ Pad\'e is used). We used the PDFs subfamily given by Eqs.~(\ref{equation: example distributions 1}), (\ref{equation: example distributions 2}), and (\ref{equation: example distributions 3}), for (a) $\nu=2$, (b) $\nu=1/2$, (c) $\nu=5$, and (d) $\nu=1/5$.}
\label{figure: delta an}
\end{figure}

One might think that one needs to know the exact underlying distribution to generate satisfactory approximations to $\aN$ and $\bN$, but this is not the major stumbling block. Let us consider for the present the family of distributions with the large-$\chi$ asymptotic behavior
\begin{equation}
\label{equation: extended exponential asymptotics simple}
	1 - F(\chi) \simeq \frac{e^{-C\chi^{\nu}}}{\chi^{\theta}} D_0 ,
\end{equation}
with $\nu,C,D_0>0$ and $\theta \in \mathbb{R}$, which is a fairly general form of $F(\chi)$, that nevertheless keeps the expressions manageable. This includes the stretched (for $0<\nu<1$) and compressed (for $\nu>1$) exponential distributions, and in particular, the Gaussian and Gamma distributions. Note that knowing the values of $\nu$, $C$, $D_0$, and $\theta$ in Eq.~(\ref{equation: extended exponential asymptotics simple}) is a minimal requirement needed to make our theory presented in Secs.~\ref{section: uncorrected}-\ref{section: tails} usable. In Sec.~\ref{section: practical} we present a method for which no knowledge of the asymptotic form of the underlying CDF is required. Working to sub-sub-leading order, evaluating Eq.~(\ref{equation: normalization sequences definition}) using Eqs.~(\ref{equation: l definition}) and (\ref{equation: extended exponential asymptotics simple}) yields
\begin{align}
\label{equation: sequences solved simple}
	&\aN \simeq \aN^{\rm s} \equiv \left[\frac{\ln(N)}{C}\right]^{1/\nu} \left\{ 1 - \frac{\theta\ln[\ln(N)]}{\nu^2\ln(N)} + \frac{\ln(D_0C^{\theta/\nu})}{\nu\ln(N)}\right\} , \nonumber \\
	&\bN \simeq \bN^{\rm s} \equiv \frac{1}{\nu C^{1/\nu} [\ln(N)]^{1-1/\nu}} \left\{ 1 - \frac{(1-\nu)\theta\ln[\ln(N)]}{\nu^2\ln(N)} + \frac{(1-\nu)\ln(D_0C^{\theta/\nu})-\theta}{\nu\ln(N)} \right\} ,
\end{align}
where the superscript ``s" stands for the standard approach. These coincide with the known formulas for the Gaussian case found in Ref.~\cite{Hall}, and with the leading order result given in Ref.~\cite{Majumdar}. Note that the $O[1/\ln(N)]$ correction to $\aN$ is necessary to satisfy the criterion on $\aN$ Eq.~(\ref{equation: conditions on sequences}), whereas the correction to $\bN$ is not needed to satisfy the corresponding demand on $\bN$.

To see how well these equations work in practice, we test them for a particular subfamily of distributions satisfying Eq.~(\ref{equation: extended exponential asymptotics simple}), with a PDF given by
\begin{equation}
\label{equation: example distributions 1}
	f(\chi) = \frac{\nu}{2\Gamma(1/\nu)} \sqrt{\frac{\Gamma(3/\nu)}{\Gamma(1/\nu)}} \exp\left\{-\left[\frac{\Gamma(3/\nu)}{\Gamma(1/\nu)}\right]^{\nu/2}|\chi|^{\nu}\right\}
\end{equation}
over the domain $\chi\in(-\infty,\infty)$, where the parameters governing the large $\chi$ asymptotics are
\begin{equation}
\label{equation: example distributions 2}
	\theta=\nu-1 , \quad C = \left[\frac{\Gamma(3/\nu)}{\Gamma(1/\nu)}\right]^{\nu/2} , \quad D_0=\frac{C^{(1-\nu)/\nu}}{2\Gamma(1/\nu)} ,
\end{equation}
and $\Gamma(\cdot)$ is the gamma function. Note that this subfamily has zero mean and unit standard deviation. In particular, $\nu=2$ is the standard Gaussian. In Fig.~\ref{figure: delta an}, we present the scaled error in $\aN$, $\delta^{\rm s} \equiv |\aN - \aN^{\rm s}|/\bN$, for the cases (a) $\nu=2$ (standard Gaussian, compressed exponential), (b) $\nu=1/2$ (stretched exponential), (c) $\nu=5$ (super-compressed exponential), and (d) $\nu=1/5$ (super-stretched exponential). Note that $\aN$ and $\bN$ denote the ``exact" values satisfying Eq.~(\ref{equation: normalization sequences definition}). We see that for (b), (c), and (d), the error $\delta^{\rm s}$ (green circles) remains above $10\%$ even for $N$s as large as $10^6$. In fact, the error does not fall below $10\%$ until $N \approx 10^{60}$ for (b), $N \approx 10^{43}$ for (c), and $N\approx 10^{12500}$ for (d). This unfortunate situation is true for other $\nu$s as well, and keeps deteriorating the further one is from $\nu=1$.

The way out of this dilemma is actually quite simple. One can directly solve the approximate equation
\begin{equation}
\label{equation: normalization sequences zero order def}
    \frac{e^{-C\alphaN^{\nu}}}{\alphaN^{\theta}} D_0 = \frac{1}{N} ,
\end{equation}
which replaces the exact $L(\chi)$ in Eq.~(\ref{equation: normalization sequences definition}) by its leading-order large-$\chi$ approximation. The solution, which we denoted above by $\alphaN$, can be expressed in terms of the Lambert W-function which obeys $\text{W}(\eta)\exp[\text{W}(\eta)]=\eta$, giving
\begin{equation}
\label{equation: normalization sequences zero order}
	\alphaN \equiv \left\{
	\begin{aligned}
		&\left\{ \frac{\theta}{\nu C} \text{W}_0 \left[ \frac{\nu C}{\theta} \left( D_0N \right)^{\nu/\theta} \right] \right\}^{1/\nu} & \theta>0 \\
		&\left[ \frac{1}{C} \ln\left(D_0N\right) \right]^{1/\nu} & \theta=0 \\
		&\left\{ \frac{\theta}{\nu C} \text{W}_{-1} \left[ \frac{\nu C}{\theta} \left( D_0N \right)^{\nu/\theta} \right] \right\}^{1/\nu} & \theta < 0
	\end{aligned} \right. .
\end{equation}
Here, $\text{W}_0(\cdot)$ is the Lambert W-function's primary real branch, which has an asymptotic expansion for $\eta\to\infty$ given by $\text{W}_0(\eta)\sim\ln(\eta)-\ln[\ln(\eta)]$, whereas $\text{W}_{-1}(\cdot)$ is the Lambert W-function's secondary real branch, which is defined on the interval $[-1/e,0)$ and has an asymptotic expansion for $\eta\to 0^-$ given by $\text{W}_{-1}(\eta)\sim\ln(-\eta)-\ln[-\ln(-\eta)]$~\cite{DLMF}. By virtue of this asymptotic behavior, $\aN^{\rm s}$ as given in Eq.~(\ref{equation: sequences solved simple}) can be retrieved from Eq.~(\ref{equation: normalization sequences zero order}). The advantage of this formula is clear, as the entire $N$-dependence is encapsulated in the single parameter $\alphaN$. This ``Lambert" approximation for $\aN$ performs much better than $\aN^{\rm s}$, as can be seen in Fig.~\ref{figure: delta an} (blue disks), where the Lambert error, $\delta^{\rm L} \equiv |\aN - \alphaN|/\bN$, is plotted together with $\delta^{\rm s}$. We see that for (b), $\delta^{\rm L}$ falls below $10\%$ already at $N\approx2800$, an improvement of roughly $57$ orders of magnitude in the range of $N$ where the approximation is useful. Similarly for $\nu=5$, $\delta^{\rm L}$ falls below $10\%$ for $N \approx 36000$.
\begin{figure}
	\includegraphics[width=1.0\textwidth]{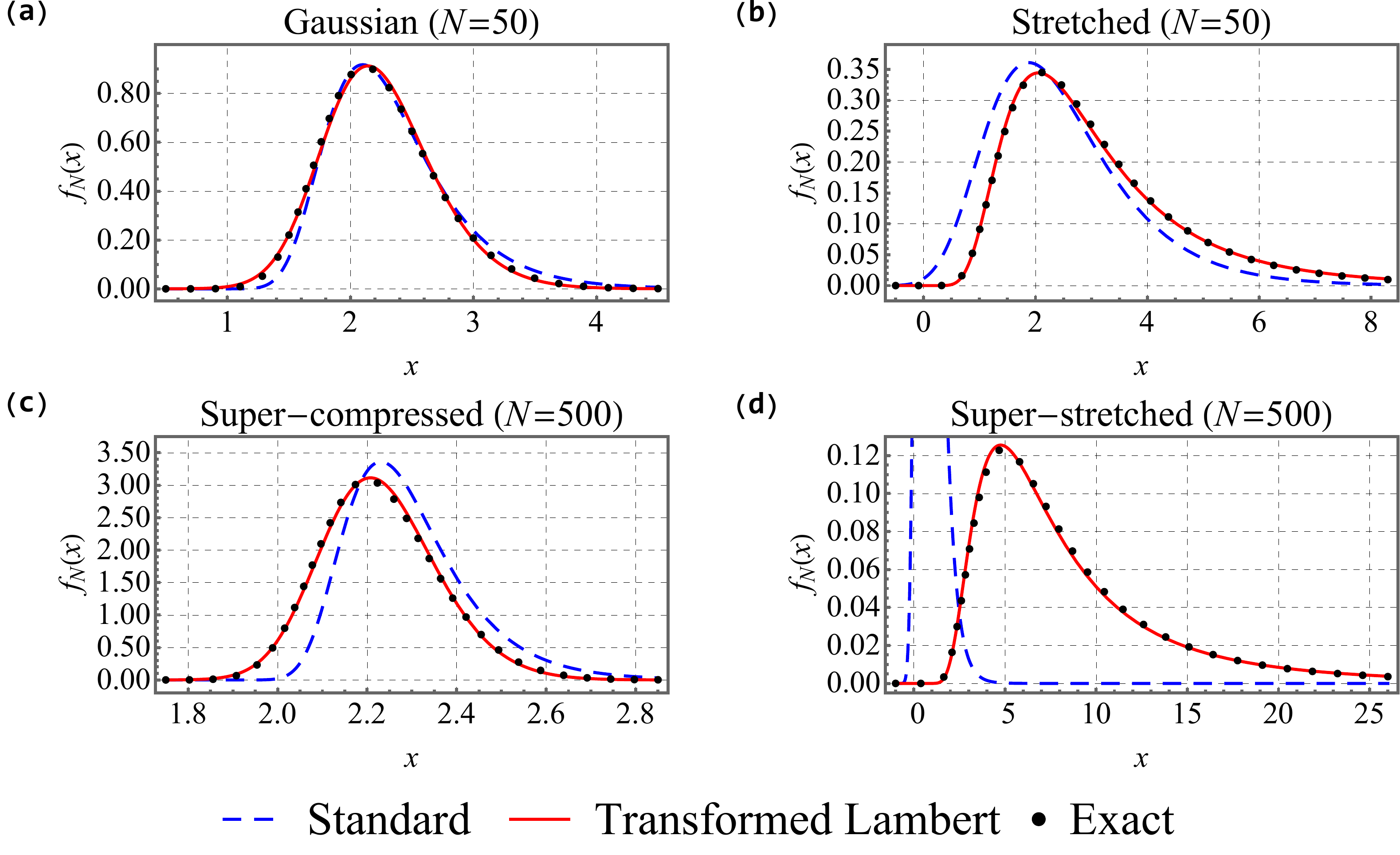}
	\caption{The PDF of the maximal value $x$ of (a,b) $N=50$ and (c,d) $N=500$ IID random variables with common PDFs given by Eq.~(\ref{equation: example distributions 1}) for four values of $\nu$: (a) $2$, (b) $1/2$, (c) $5$, and (d) $1/5$. The exact values (dotted black) are compared to the two different types of approximations: the standard $\ln(N)$ expansion given by $f_N(x) \simeq (1/\bN^{\rm s}) g_{\infty}[(x-\aN^{\rm s})/\bN^{\rm s}]$ (dashed blue), and the transformed Lambert method given by Eqs.~(\ref{equation: normalization sequences expansion change variables}), (\ref{equation: normalization sequences zero order change variables}), and (\ref{equation: fn via gumbel change variables}) (solid red). For the latter and in the case of $\nu>1$, a $[1/2]$ Pad\'e approximants in the variable $(\alphaN^w)^{-1}$ are implied for Eq.~(\ref{equation: normalization sequences expansion change variables}). The transformed Lambert method clearly holds very well already for intermediate $N$s. As stated, an underlying distribution which is far from a pure exponential of $\nu=1$ has a slower convergence rate, see Fig.~\ref{figure: gumbel zero larger} in appendix \ref{section: larger} for a replot using larger $N$s.}
\label{figure: gumbel zero}
\end{figure}

To improve the quality of our approximation for $\aN$ yet further, and widen the range of $N$s we can treat, we must utilize more knowledge of the asymptotic behavior of $F(\chi)$. For example, if we assume the asymptotic expansion has the form
\begin{equation}
\label{equation: extended exponential asymptotics}
	1 - F(\chi) \simeq \frac{e^{-C\chi^{\nu}}}{\chi^{\theta}}D_0 \left(1+\frac{D_1}{\chi^{\nu}}+\frac{D_2}{\chi^{2\nu}}\right) ,
\end{equation}
where for our example family of distributions
\begin{equation}
\label{equation: example distributions 3}
    D_1 = -\left(1-\frac{1}{\nu}\right)\frac{1}{C} , \quad D_2 = \left(1-\frac{1}{\nu}\right)\left(2-\frac{1}{\nu}\right)\frac{1}{C^2} ,
\end{equation}
then we can make additional progress. The key here is to express the expansion not in terms of $N$, but in terms of $\alphaN$, our zeroth order Lambert approximation for $\aN$. We can similarly express $\bN$ as well in terms of $\alphaN$. We find to order $1/\alphaN^{3\nu}$,
\begin{align}
\label{equation: normalization sequences expansion}
	&\aN \simeq \aN^{\rm L,3} \equiv \alphaN \left[1+\frac{D_1}{\nu C\alphaN^{2\nu}}-\frac{2\theta D_1 + \nu C (D_1^2-2D_2)}{2\nu^2C^2\alphaN^{3\nu}}\right] , \\
	&\bN \simeq \bN^{\rm L,3} \equiv \frac{1}{\nu C\alphaN^{\nu-1}} \left[1-\frac{\theta}{\nu C\alphaN^{\nu}} + \frac{\theta^2-\nu(2\nu-1) CD_1}{\nu^2C^2\alphaN^{2\nu}} - \frac{2\theta^3-2\nu\theta(5\nu-2)CD_1-\nu^2(3\nu-1)C^2(D_1^2-2D_2)}{2\nu^3C^3\alphaN^{3\nu}}\right] , \nonumber
\end{align}
These expansions have the added advantage over the standard approximation, in addition to the higher accuracy of the zeroth-order term, that they are standard power series in $\alphaN^{-\nu}$, with no iterated logarithm terms. This means that if needed, one can use techniques such as Pad\'e approximants to help accelerate the convergence rate. We find that for the compressed cases of $\nu>1$, the $[1/2]$ Pad\'e approximants of the sequences in Eq.~(\ref{equation: normalization sequences expansion}) perform better than the regular power series, whereas for the stretched cases of $0<\nu<1$, it is better to use the series expansions as expressed above. The scaled error $\delta^{\rm L,3}=|\aN-\aN^{\rm L,3}|/\bN$ is also indicated in Fig.~\ref{figure: delta an} (red triangles, red squares for Pad\'e form), where we see a drastic improvement in the scaled error for all cases.

\subsection{Changing variables}

As we saw, at leading order the EV distribution can be approximated by the Gumbel distribution characterized by the two parameters, $\aN$ and $\bN$. However, the further $\nu$ departs from unity, the more the shape of the distribution deviates from Gumbel. This is related to the fact that as $\nu \to 0$, the distribution acquires a fat tail and the Gumbel description breaks down, with the scaling limit being a Fr\'echet distribution. Similarly, as $\nu\to\infty$, the distribution becomes compact, with a Weibull scaling limit. In other words, this situation occurs for common distributions that have an $L$ which is far from a linear function, causing in turn the Taylor approximation Eq.~(\ref{equation: l expansion}) to fail. This problem can be seen in Fig.~\ref{figure: gumbel zero}, where not only is the peak location poorly given by the standard approximation for all but the Gaussian case, but the shape is distinctly different from that of the Gumbel distribution in the non-Gaussian cases.

A simple remedy for this problem is given by the expedient of changing variables as $\omega\sim\chi^{\nu}$ for $\chi\to\infty$, in terms of which the underlying distribution has a simple exponential falloff as its dominant behavior. Consequently, the EV distribution for $w \sim x^{\nu}$ and $x\to\infty$ is thus well-described by a Gumbel distribution, with parameters $\aN^w = \aN^\nu$ and $\bN^w = \nu\aN^{\nu-1}\bN$, namely
\begin{align}
\label{equation: normalization sequences expansion change variables}
	&\aN^w \simeq \alphaN^w \left[1+\frac{D_1}{C(\alphaN^w)^2}-\frac{2\theta_wD_1 + C(D_1^2-2D_2)}{2C^2(\alphaN^w)^3}\right] , \nonumber \\
	&\bN^w \simeq \frac{1}{C} \left[1-\frac{\theta_w}{C\alphaN^w} + \frac{\theta_w^2-CD_1}{C^2(\alphaN^w)^2} - \frac{2\theta_w^3-6\theta_wCD_1-2C^2(D_1^2-2D_2)}{2C^3(\alphaN^w)^3}\right] ,
\end{align}
where $\theta_w \equiv \theta/\nu$, and
\begin{equation}
\label{equation: normalization sequences zero order change variables}
	\alphaN^w \equiv \left\{
	\begin{aligned}
		& \frac{\theta_w}{C} \text{W}_0 \left[ \frac{C}{\theta_w} \left( D_0N \right)^{1/\theta_w} \right] & \theta_w>0 \\
		& \frac{1}{C} \ln\left(D_0N\right) & \theta_w=0 \\
		& \frac{\theta_w}{C} \text{W}_{-1} \left[ \frac{C}{\theta_w} \left( D_0N \right)^{1/\theta_w} \right] & \theta_w < 0
	\end{aligned} \right. .
\end{equation}
Note that the scaled error of $\aN^w$ is equal to that of $\aN$ to leading order, hence our previous work in approximating $\aN$ directly carries over. The Gumbel distribution in $w$ translates directly to our new PDF for $x$,
\begin{equation}
\label{equation: fn via gumbel change variables}
	f_N(x) = \nu x^{\nu-1} f_N^w\left(x^{\nu}\right) = \frac{\nu x^{\nu-1}}{\bN^w} g_N^w\left(\frac{x^{\nu}-\aN^w}{\bN^w}\right) \simeq \frac{\nu x^{\nu-1}}{\bN^w} g_{\infty}\left(\frac{x^{\nu}-\aN^w}{\bN^w}\right) .
\end{equation}
Figure~\ref{figure: gumbel zero} shows the EV PDFs $f_N(x)$ for the four examples stated above. The exact values are compared to the standard Gumbel approximation given by $f_N(x) \simeq (1/\bN^{\rm s}) g_{\infty}[(x-\aN^{\rm s})/\bN^{\rm s}]$, and to our transformed Lambert approximation given by Eqs.~(\ref{equation: normalization sequences expansion change variables}), (\ref{equation: normalization sequences zero order change variables}), and (\ref{equation: fn via gumbel change variables}). As with Eq.~(\ref{equation: normalization sequences expansion}), a $[1/2]$ Pad\'e approximant in the variable $(\alphaN^w)^{-1}$ was employed to Eq.~(\ref{equation: normalization sequences expansion change variables}) for the compressed cases. We changed variables according to $w = \text{sign}(x)|x|^{\nu}$, which is consistent with the asymptotics $w \sim x^{\nu}$ described above. The combined usage of the Lambert scaling and the variable transformation excellently match the exact results, without applying any corrections to the Gumbel distribution. In appendix \ref{section: larger}, we replot all panels with an $N$ that is larger by a factor of $10^3$, see Fig.~\ref{figure: gumbel zero larger}, demonstrating the slow rate of convergence for the standard approximation. Note that none of the two methods discussed above can perform as well alone, hence they are complementary.
\begin{figure}
	\includegraphics[width=1.0\textwidth]{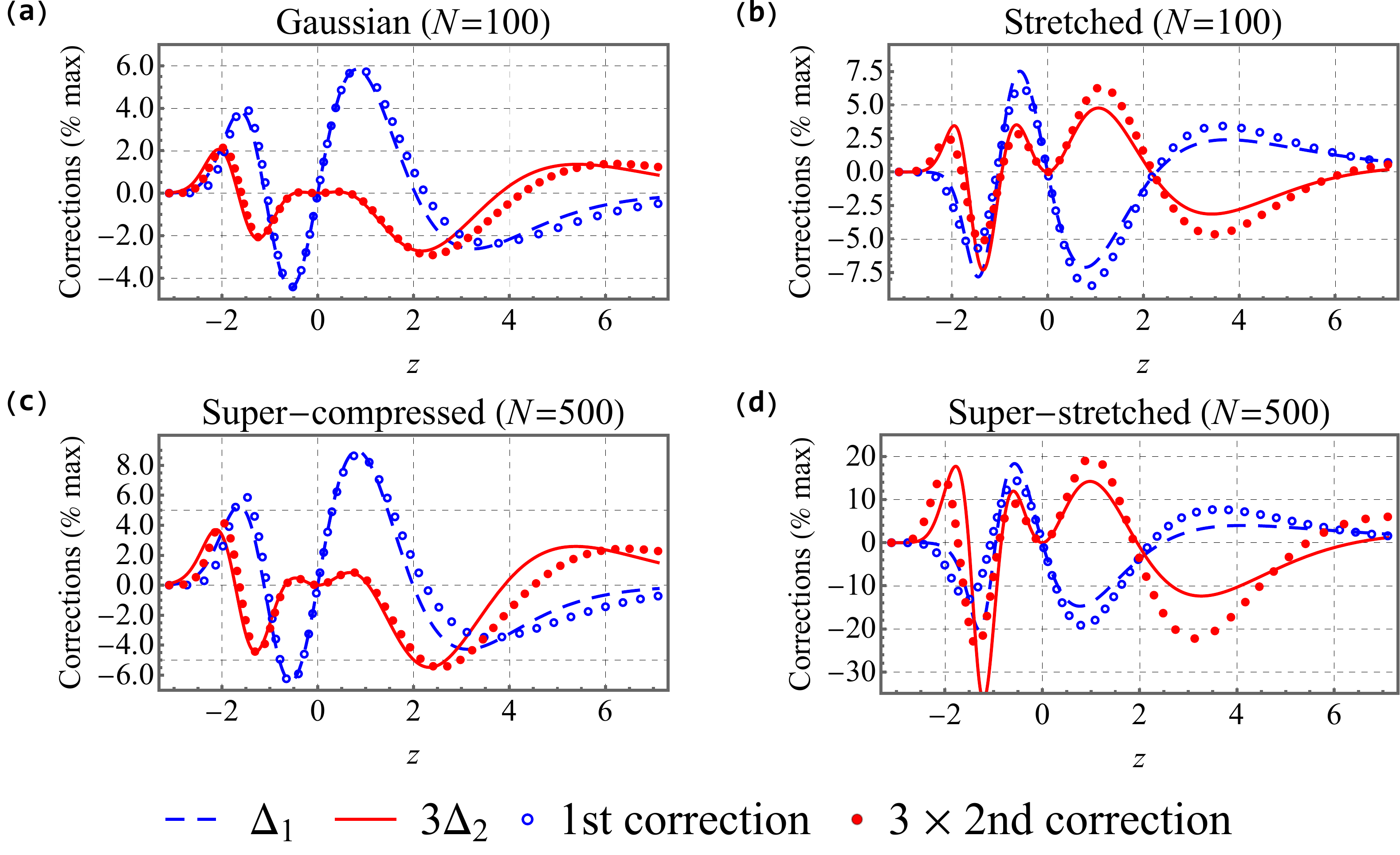}
	\caption{The first and second corrections to the Gumbel approximation of the EV PDFs, relative to the maximum value $1/e$ of the Gumbel distribution, for the four test distributions of Eq.~(\ref{equation: example distributions 1}) with $\nu=$ (a) $2$, (b) $1/2$, (c) $5$, and (d) $1/5$, where $N=$ (a,b) $100$ and (c,d) $500$. The first correction $\Delta_1$ (dashed blue) defined in Eq.~(\ref{equation: delta 1 definition}) has a magnitude of $\approx 10\%$ with respect to the maximal value of $g_{\infty}(z)$, $1/e$. It follows well its predicted shape (blue circles) seen on the right hand side of Eq.~(\ref{equation: delta 1 definition}). The second correction $\Delta_2$ (solid red), defined in Eq.~(\ref{equation: delta 2 definition}), is multiplied by $3$ for visibility. It follows its predicted shape (red disks) seen on the right hand side of Eq.~(\ref{equation: delta 2 definition}), and has a smaller magnitude than the first-order correction. Here we used the exact values of $\aN$, $\bN$, $c_2$, and $c_3$.}
\label{figure: corrections delta}
\end{figure}
\begin{figure}
	\includegraphics[width=1.0\textwidth]{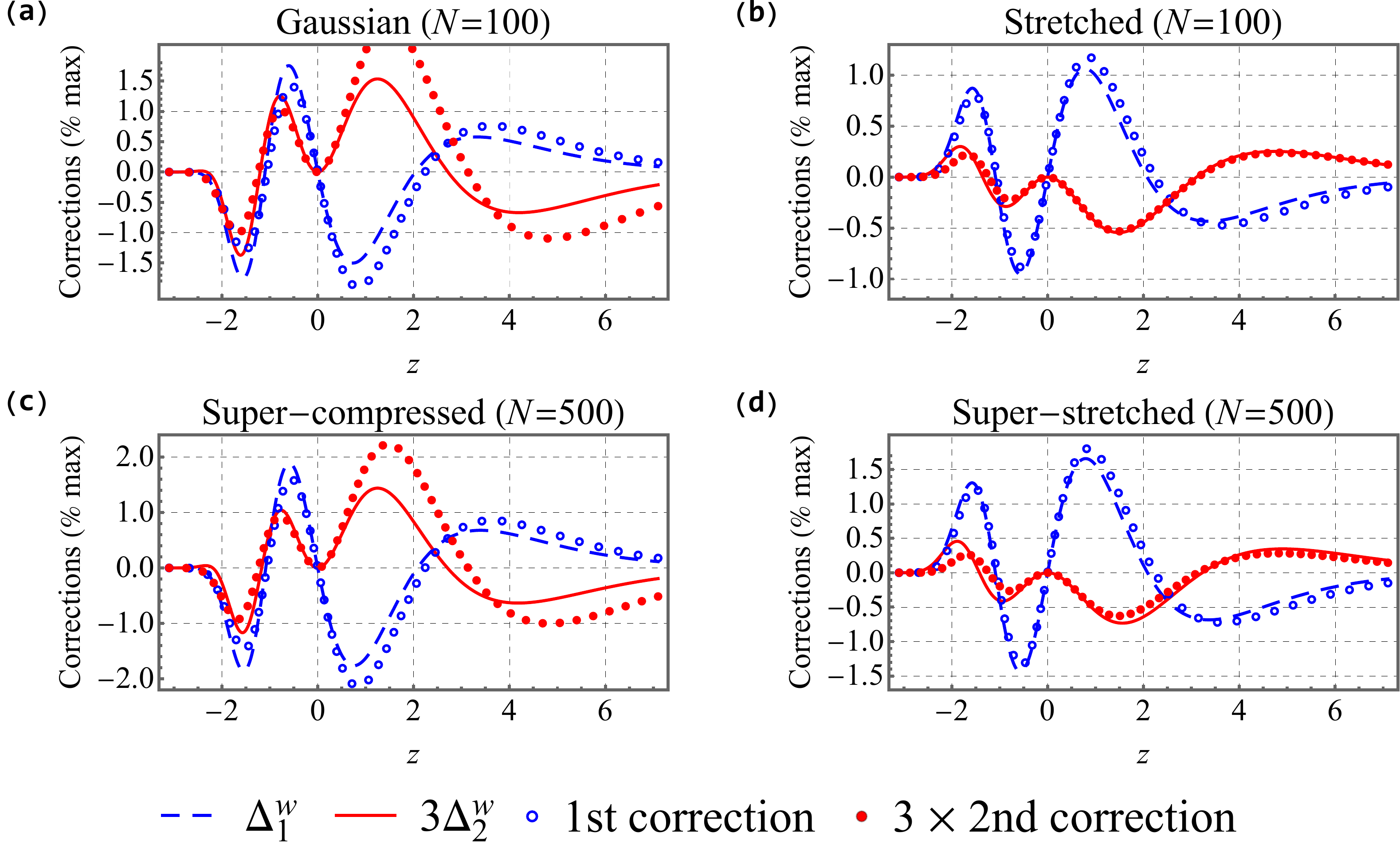}
	\caption{The first and second corrections to the transformed Gumbel approximation of the EV PDFs, relative to the maximum value $1/e$ of the Gumbel distribution, for the four test distributions of Eq.~(\ref{equation: example distributions 1}) with $\nu=$ (a) $2$, (b) $1/2$, (c) $5$, and (d) $1/5$, where $N=$ (a,b) $100$ and (c,d) $500$. The first correction $\Delta_1^w$ (dashed blue) defined in Eq.~(\ref{equation: delta 1 definition}) has a magnitude of $\approx 1\%$ with respect to the maximal value of $g_{\infty}(z)$, $1/e$. It follows well its predicted shape (blue circles) seen on the right hand side of Eq.~(\ref{equation: delta 1 definition}). The second correction $\Delta_2^w$ (solid red), defined in Eq.~(\ref{equation: delta 2 definition}), is multiplied by $3$ for visibility. It follows its predicted shape (red disks) seen on the right hand side of Eq.~(\ref{equation: delta 2 definition}), and has a smaller magnitude than the first-order correction. We used the exact values of $\aN^w$, $\bN^w$, $c_2^w$, and $c_3^w$. The magnitude of the corrections after performing the change of variables is noticeably smaller for all cases.}
\label{figure: corrections change}
\end{figure}

\section{Corrections to the Gumbel distribution}
\label{section: corrections}

We now consider corrections to the Gumbel distribution itself. Let us continue from Eqs.~(\ref{equation: l expansion}) and (\ref{equation: normalization sequences definition}) by taking one additional term from the expansion of $L$. In what follows, we suppress the argument $\aN$ of $c_n(\aN)$. We obtain the Gumbel distribution to linear order along with the first correction in $\aN$,
\begin{equation}
\label{equation: general first correction}
	G_N(z) \simeq G_{\infty}(z) \left[ 1 + c_2\frac{z^2}{2}e^{-z} \right] ,
\end{equation}
which leads to the approximate PDF
\begin{equation}
\label{equation: general first correction pdf}
    g_N(z) \simeq g_{\infty}(z) \left[ 1 + c_2\frac{z}{2}\left(2-z+e^{-z}z\right) \right] .
\end{equation}
This first order correction is already known from the renormalization-group works by Gy\"{o}rgyi, et al. \cite{Gyorgyi1,Gyorgyi2,Gyorgyi3}. Indeed, we see that it has a universal functional shape, while the numerical prefactor $c_2$ depends on the specifics of the underlying distribution $F$. The second order correction relies on the additional numerical parameter $c_3$. In the renormalization-group language, each additional term comes from a subdominant eigenvalue of the renormalization operator, but here the procedure is simply a Taylor expansion of the appropriate function, namely $L$. Using Eqs.~(\ref{equation: normalization sequences definition}) and (\ref{equation: extended exponential asymptotics simple}), one can show that $c_2 \sim (\nu-1)/[\nu\ln(N)]$ and that $c_2^w = c_2 + (1-\nu)\bN/\aN \sim -\theta_w/[\ln(N)]^2$, which occurs as the transformation of variables gives an effective $\nu=1$. Hence, the transformed coefficient $c_2^w$ is down by an additional factor of $1/\ln(N)$. In order to illustrate this first correction, we define
\begin{align}
\label{equation: delta 1 definition}
	\Delta_1 &\equiv \frac{1}{1/e}\left[\bN f_N\left( \aN + \bN z \right) - g_{\infty}(z) \vphantom{\frac{1}{1}}\right] \simeq \frac{c_2}{1/e} g_{\infty}(z) \frac{z}{2}\left(2-z+e^{-z}z \vphantom{\frac{1}{1}}\right) , \nonumber \\
	\Delta_1^w &\equiv \frac{1}{1/e}\left[\bN^w f_N^w\left( \aN^w + \bN^w z \right) - g_{\infty}(z) \vphantom{\frac{1}{1}}\right] \simeq \frac{c_2^w}{1/e} g_{\infty}(z) \frac{z}{2}\left(2-z+e^{-z}z \vphantom{\frac{1}{1}}\right) .
\end{align}
These are the differences between the exact EV PDF for the scaled variable $z$ and the Gumbel approximation, normalized to the maximal value of $g_{\infty}(z)$, $1/e$, where the superscript $w$ denotes the variable change $w = \text{sign}(x)|x|^{\nu}$. Figures~\ref{figure: corrections delta} and \ref{figure: corrections change} show $\Delta_1$ and $\Delta_1^w$ (dashed blue), respectively, for our four examples, together with the predicted shapes of the first correction as given in Eq.~(\ref{equation: delta 1 definition})'s right hand side (blue circles). The differences follow the predicted curves well, and one can see that the relative magnitude of the first correction significantly reduces when applying the variable change, from $\sim 10\%$ to $\sim 1\%$.
\begin{figure}
	\includegraphics[width=1.0\textwidth]{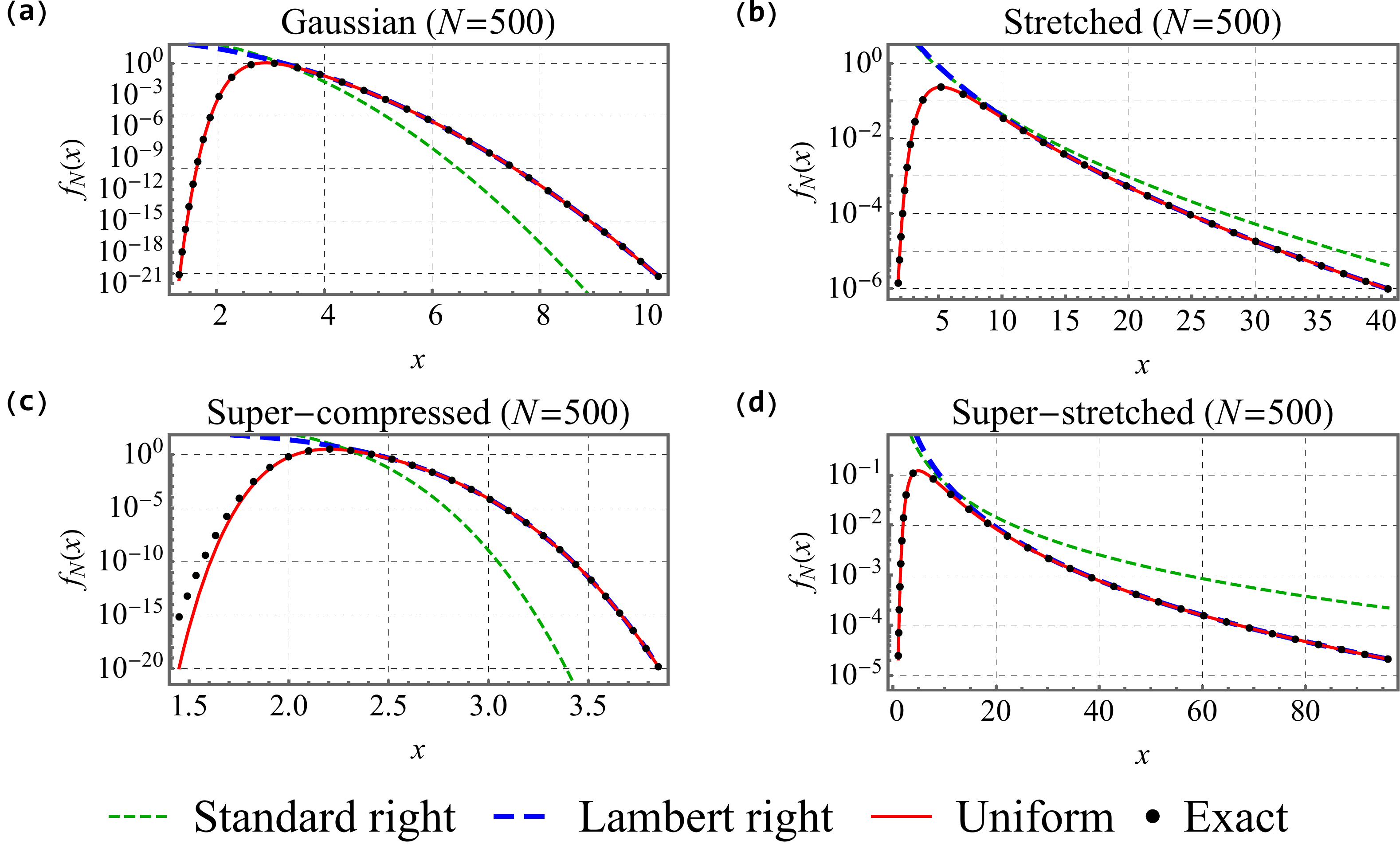}
	\caption{The left and right tails of the PDF of the maximal value $x$ of $N=500$ IID random variables with common PDFs given by Eq.~(\ref{equation: example distributions 1}) for four values of $\nu$: (a) $2$, (b) $1/2$, (c) $5$, and (d) $1/5$. The exact values (dotted black) excellently match our Lambert scaled approximation for the right tail (thick dashed blue), given by Eqs.~(\ref{equation: normalization sequences zero order change variables}) and (\ref{equation: large deviations principle extended}) in the relevant regime. Also seen are the large deviation results of Ref.~\cite{Rita} for the right tail, given by Eq.~(\ref{equation: large deviations principle simple}) (short-dashed green). The uniform approximation of the EV PDF (solid red), given by Eqs.~(\ref{equation: normalization sequences zero order change variables}) and (\ref{equation: extended uniform approximation solution}), nicely match with the exact values for all relevant $x$s.}
\label{figure: tails approx}
\end{figure}

Next, we demonstrate the second-order correction, going beyond the renormalization-group calculations of Refs.~\cite{Gyorgyi1,Gyorgyi2,Gyorgyi3}. One can show that $c_3 \sim (\nu-1)(\nu-2)/[\nu\ln(N)]^2$ and that $c_3^w = c_3 - 3(\nu-1)c_2\bN/\aN + (\nu-1)(2\nu-1)(\bN/\aN)^2 \sim 2\theta_w/[\ln(N)]^3$, which occurs for the same reason as before. Thus, $c_3^w$ is down by an additional factor of $1/\ln(N)$, and the second-order correction to the EV PDF will be different if one applies this change of variables. For the original variable, extracting yet another term from Eq.~(\ref{equation: l expansion}), we arrive at the approximate CDF
\begin{equation}
\label{equation: general second correction}
	G_N(z) \simeq G_{\infty}(z) \left[1 + c_2\frac{z^2}{2}e^{-z} + c_3\frac{z^3}{6}e^{-z} - c_2^2\frac{z^4}{8}e^{-z}\left(1-e^{-z}\right) \right] ,
\end{equation}
with an approximate PDF of
\begin{equation}
\label{equation: general second correction pdf}
    g_N(z) \simeq g_{\infty}(z) \left\{1 + c_2\frac{z}{2}\left(2-z+e^{-z}z\right) + c_3\frac{z^2}{6}\left(3-z+e^{-z}z\right) - c_2^2\frac{z^3}{8} \left[4-z-e^{-z}\left(4-3z+e^{-z}z\right)\right] \right\} .
\end{equation}
Note that in the language of the transformed variable $w$, the term proportional to $c_2^2$ is of a higher order. Hence, in this representation, the second-order correction is also of universal behavior. In order to illustrate this correction, we define
\begin{align}
\label{equation: delta 2 definition}
	\Delta_2 &\equiv \frac{1}{1/e}\left\{\bN f_N\left( \aN + \bN z \right) - g_{\infty}(z)\left[ 1 + c_2\frac{z}{2}\left(2-z+e^{-z}z\right) \right] \right\} \nonumber \\
	&\simeq \frac{g_{\infty}(z)}{1/e} \left\{ c_3\frac{z^2}{6}\left(3-z+e^{-z}z\right) - c_2^2\frac{z^3}{8} \left[4-z-e^{-z}\left(4-3z+e^{-z}z\right)\right] \right\} , \nonumber \\
	\Delta_2^w &\equiv \frac{1}{1/e}\left\{\bN^w f_N^w\left( \aN^w + \bN^w z \right) - g_{\infty}(z) \left[ 1 + c_2^w\frac{z}{2}\left(2-z+e^{-z}z\right) \right] \right\} \simeq \frac{c_3^w}{1/e} g_{\infty}(z) \frac{z^2}{6}\left(3-z+e^{-z}z \vphantom{\frac{1}{1}}\right) .
\end{align}
These are the differences between the exact EV PDF and the first order Gumbel approximation in the $z$ coordinate, normalized to the maximal value of $g_{\infty}(z)$, $1/e$, where the superscript $w$ denotes the variable change $w = \text{sign}(x)|x|^{\nu}$. Figures~\ref{figure: corrections delta} and \ref{figure: corrections change} show $\Delta_2$ and $\Delta_2^w$ (solid red), respectively, for our four examples, together with the predicted shapes of the second correction as given in Eq.~(\ref{equation: delta 2 definition})'s right hand side (red disks). The differences follow the predicted curves, and one can see that the relative magnitude of the second correction significantly reduces when changing variables.

We conclude this section with a calculation of the EV distribution's moments, which are given by
\begin{equation}
\label{equation: moments definition}
	\left< x^m \right> \equiv \int_{-\infty}^{\infty}\text{d}x \, f_N(x) x^m .
\end{equation}
As done above, we change variables to $w=|x|^{\nu}\text{sign}(x)$, with an inverse of $x=|w|^{1/\nu}\text{sign}(w)$. Then, Eq.~(\ref{equation: moments definition}) becomes
\begin{align}
\label{equation: moments expansion variable change}
	\left< x^m \right> \simeq (\aN^w)^{m_w} \sum_{n=0}^{\infty} \frac{\Gamma(n-m_w)}{n!\Gamma(-m_w)} \left(-\frac{\bN^w}{\aN^w}\right)^n \int_{-\infty}^{\infty}\text{d}z \, g_N^w(z) z^n ,
\end{align}
up to exponentially small corrections, where $m_w \equiv m/\nu$. Integrating the PDF Eq.~(\ref{equation: general first correction pdf}) and plugging in the expansions for $\aN^w$, $\bN^w$, and $c_2^w$ (not shown) gives for the $m$th moment
\begin{equation}
\label{equation: moments expansion}
	\left< x^m \right> \simeq (\alphaN^w)^{m_w} \left[ 1 + \frac{m_w\gamma}{C\alphaN^w} + \frac{m_w(m_w-1)(6\gamma^2+\pi^2)-12m_w\theta_w\gamma +12m_wCD_1}{12C^2(\alphaN^w)^2} \right] ,
\end{equation}
where $\gamma\approx 0.5772$ is the Euler–Mascheroni constant. An important advantage of our series expansion is that it allows one to obtain higher-order corrections to Eq.~(\ref{equation: general first correction pdf}) rather easily, see e.g. Eq.~(\ref{equation: general second correction pdf}), hence Eq.~(\ref{equation: moments expansion}) can be extended to arbitrary orders.
\begin{figure}
	\includegraphics[width=1.0\textwidth]{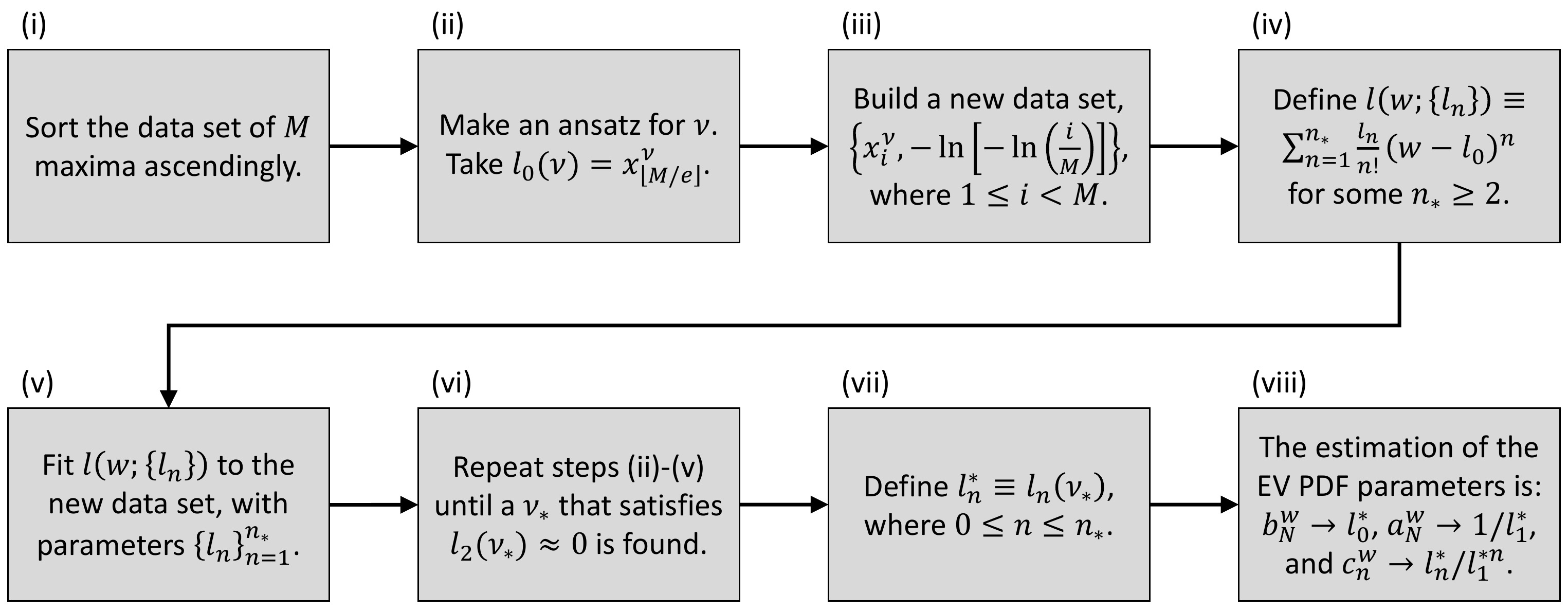}
	\caption{A flowchart describing a suggested algorithm for fitting numerical data set of $M$ maxima to our theory. In step (ii), $\lfloor\cdot\rfloor$ denotes the $\text{floor}(\cdot)$ function. In step (iv), the value of $n_*$ determines the highest order correction term of the Gumbel distribution to be obtained. In step (vi), the value of $\nu_*$ stands for the optimal variable change exponent, and generally does not have to be an integer. In step (viii), the referred EV PDF is given by Eq.~(\ref{equation: fn via gumbel change variables}). Note that in all of the considered examples, the $\{c_n^w\}$ parameters were not needed to yield a good match, see Figs.~\ref{figure: gumbel fit} and \ref{figure: minimum case}(d).}
\label{figure: gumbel chart}
\end{figure}

\section{The far tails}
\label{section: tails}

We now turn to discuss the far tails. In the far right tail, $L(\chi)$ is no longer well-approximated by its expansion around $\aN$, and so universality breaks down. In this regime, $F(\chi)$ is exponentially close to $1$, and as such one can always write $F_N(x) \simeq 1-N[1-F(x)]$, with exponentially small corrections. Exploiting the asymptotics Eq.~(\ref{equation: extended exponential asymptotics simple}) and reexpressing $N$ using $\alphaN$ via Eq.~(\ref{equation: normalization sequences zero order def}), we have $F_N(x) \simeq 1-(\alphaN/x)^{\theta}\exp[-C(x^{\nu}-\alphaN^{\nu})]$, which yields
\begin{equation}
\label{equation: large deviations principle extended}
	f_N(x) \simeq \nu Cx^{\nu-1} \left(\frac{\alphaN}{x}\right)^{\theta} e^{-C\alphaN^{\nu}\left[\left(x/\alphaN\right)^{\nu}-1\right]} .
\end{equation}
Hence, the speed and scaling are $u_N = C\alphaN^{\nu}$ and $s_N = \alphaN$, respectively, where the rate function is $\psi(\xi) = \xi^{\nu}-1$. This formula extends the large deviations approach of Gulliano and Macci \cite{Rita}, for which $1-F_N(x) \approx \exp\{-\ln(N)[(x/s_N^{\rm s})^{\nu}-1]\}$, and
\begin{equation}
\label{equation: large deviations principle simple}
	f_N(x) \approx \nu x^{\nu-1} \frac{\ln(N)}{(s_N^{\rm s})^{\nu}} e^{-\ln(N)\left[\left(x/s_N^{\rm s}\right)^{\nu}-1\right]} , \quad 1-F\left(s_N^{\rm s}\right) \equiv \frac{1}{N} ,
\end{equation}
so that the speed was $u_N^{\rm s}=\ln(N)$. Since to leading order in $N$ one has $u_N \simeq u_N^{\rm s}$ and $s_N \simeq s_N^{\rm s}$, the leading-order $N$ dependence of the two formulas is identical. However, as with the Gumbel bulk approximation, this leading order is by far too simplistic to provide accurate predictions. Figure~\ref{figure: tails approx} shows our results and the exact numerical values of the PDFs for our four cases. Even at its base level without corrections depending on $D_1$ and $D_2$, Eq.~(\ref{equation: large deviations principle extended}) is in excellent agreement to the exact values. Also presented are the large deviation results of Ref.~\cite{Rita} given by Eq.~(\ref{equation: large deviations principle simple}).

Constructing an approximation to the left tail is a matter of interest too, since the Gumbel approximation fails at both ends. It turns out that two sub-regimes exists for the left tail, corresponding to an extreme left tail where $x\to-\infty$, and to a moderate left tail for which $1\ll x\ll\langle x\rangle$. The former regime is less interesting though, as the probability to encounter such an event is extraordinary small, and thus we focus on the latter case. In this regime, $1-F(\chi)$ is still small, though much larger than $1/N$. In fact, we can still write $F_N(x) \simeq \exp\{-N[1-F(x)]\}$, however we cannot expand further. Repeating the above procedure leads to the uniform approximation, however this time we use the extended asymptotic version Eq.~(\ref{equation: extended exponential asymptotics}). Tackling the small $x$ divergence of the extra terms is done by replacing it with a $[1/1]$ Pad\'e approximant in the variable $1/x^{\nu}$. Differentiating yields the uniform approximation as
\begin{align}
\label{equation: extended uniform approximation solution}
	&F_N(x) \approx \exp\left[ -\left(\frac{\alphaN}{x}\right)^{\theta} e^{-C\left(x^{\nu}-\alphaN^{\nu}\right)} \frac{D_1x^{\nu}+D_1^2-D_2}{D_1x^{\nu}-D_2} \right] , \nonumber \\
	&f_N(x) \approx \nu C x^{\nu-1} \left(\frac{\alphaN}{x}\right)^{\theta} \exp\left[ -C\left(x^{\nu} -\alphaN^{\nu} \right) - \left(\frac{\alphaN}{x}\right)^{\theta} e^{-C\left(x^{\nu}-\alphaN^{\nu}\right)} \frac{D_1x^{\nu}+D_1^2-D_2}{D_1x^{\nu}-D_2} \right] .
\end{align}
This expression is valid for every $x$ which satisfies $1-F(x) \ll 1/\sqrt{N}$. In particular, it describes well the moderate left tail, see Fig.~\ref{figure: tails approx}. The large deviations and Gumbel forms are obtainable from Eq.~(\ref{equation: extended uniform approximation solution}) in the appropriate limits.
\begin{figure}
	\includegraphics[width=1.0\textwidth]{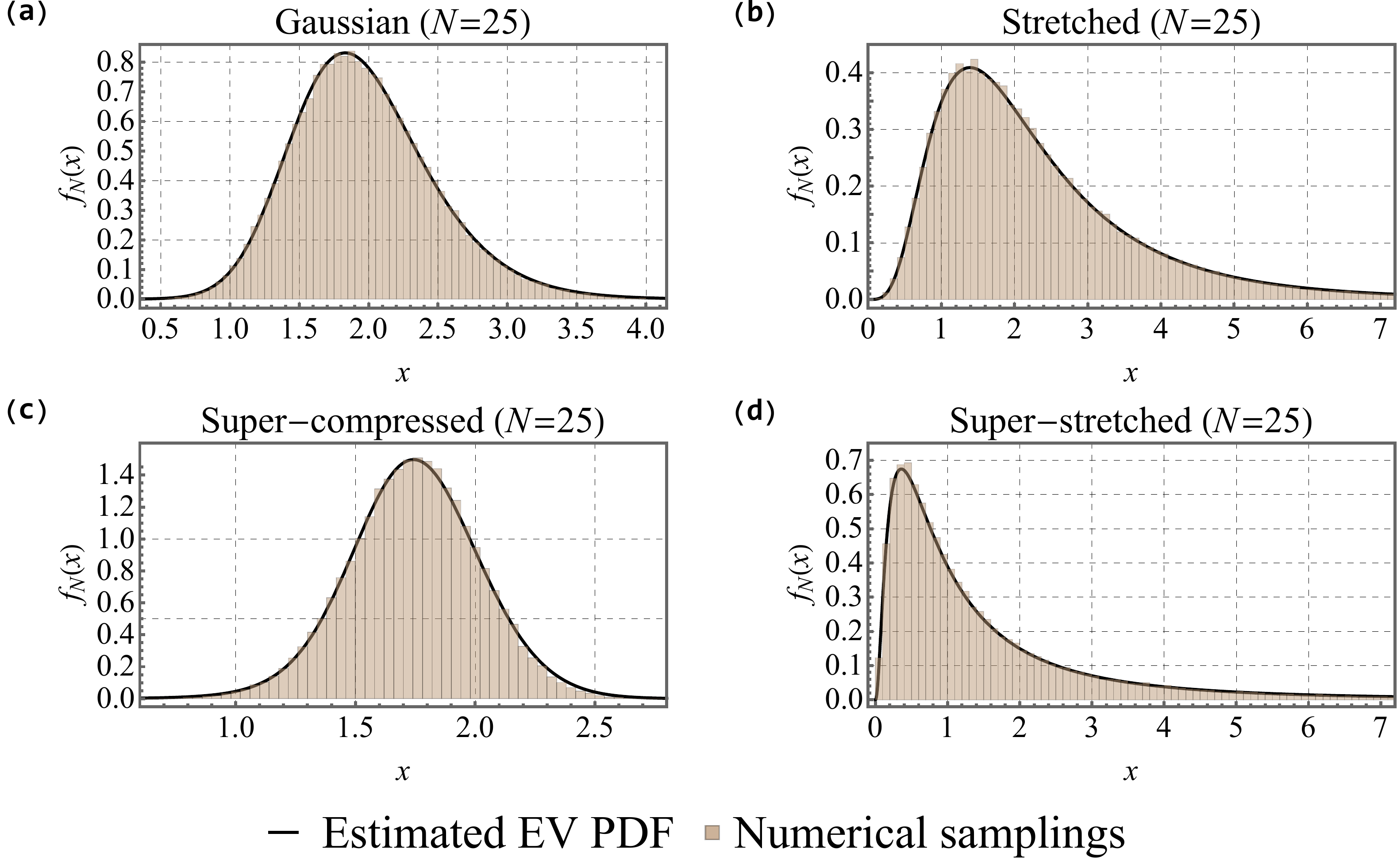}
	\caption{The PDF of the maximal value $x$ of $N=25$ IID random variables with common PDFs given by Eq.~(\ref{equation: example distributions 1}) for four values of $\nu$: (a) $2$, (b) $1/2$, (c) $5$, and (d) $1/5$. For each case we sampled $10^5$ maxima, and used these to extract an estimate for the EV PDF. Using $\nu$, $\aN^w$, and $\bN^w$ as fit parameters for the rightmost hand side of Eq.~(\ref{equation: fn via gumbel change variables}) according to the algorithm described in Fig.~\ref{figure: gumbel chart} yields an excellent match to the simulated data, without assumed knowledge of the underlying PDFs. Note that not $N$ nor $\nu$ need to be known for this procedure to work. As this algorithm renders $c_2^w \approx 0$, the plotted curves do not visibly change when adding the first correction.}
\label{figure: gumbel fit}
\end{figure}

\section{A data analysis approach}
\label{section: practical}

While above we assumed quite a general shape for the asymptotic form of the underlying distribution, in many cases one lacks knowledge of one or more of its parameters, i.e. $\nu$, $C$, $D_0$, and $\theta$. Still, this turns out to not pose a problem, as from a practical point of view, one has an excellent parameterization of the EV PDF in terms of a very small number of parameters, namely $\aN$, $\bN$, and if needed $c_2$ and $c_3$. To find these parameters given a data set of $M$ maxima $\{x_i\}$ that are assumed to follow the Gumbel limit, one can use the algorithm presented in Fig.~\ref{figure: gumbel chart}. One starts by sorting the data set ascendingly in step (i), as plotting $i/M$ as a function of $w_i = x_i^{\nu}$ for $1 \le i \le M$ essentially gives the empirical EV CDF $F_N^w(w)$. Next, by using Eq.~(\ref{equation: normalization sequences definition}) for the changed variable $w$, i.e. $\exp[L_w(\aN^w)]=N$, one has $F_N^w(\aN^w)=1/e$. This means that an estimate for $\aN^w$ can be obtained from an index which obeys $i/M \approx 1/e$, namely $i = \lfloor M/e \rfloor$, where $\lfloor\cdot\rfloor$ is the $\text{floor}(\cdot)$ function, see step (ii). Then, the quantity $-\ln[-\ln(i/M)]$ versus the argument $x_i^{\nu}$ gives the empirical value of $L_w(w)-\ln(N) = L_w(w)-L_w(\aN^w)$, which corresponds to the $w$-version of the expansion Eq.~(\ref{equation: l expansion}). Step (iv) allows for two fitting schemes. The first is truncating the data set created in step (iii) for a low-order polynomial fit around $0$ in the variable $x-l_0$. The second is fitting more of the said data set to a high-order polynomial and reading off the low-order coefficients, in which case the higher terms take care of the global behavior. Note that $n_*$ determines what is the highest-order correction term of the Gumbel distribution to be obtained. Determining the appropriate power $\nu_*$ for the variable change method is done in step (vi), by demanding that the post-transformation $L_w^{(2)}(\aN^w)$ vanish, which ensures that $L_w$ is locally quite linear near $\aN^w$. Note also that one does not need to know the values of $N$ and $\nu$ to implement this algorithm.

The results of this procedure for our four examples with $N=25$ are presented in Fig.~\ref{figure: gumbel fit}, and excellently reproduce the central region of $f_N(x)$ without any assumed knowledge of the underlying distributions. We employed high-order polynomial fits with $n_*=5$ for all cases, but only used fit parameters of the zero order, i.e. $\aN^w$ and $\bN^w$, when plotting. Note that this procedure is not intended to provide an estimation of the true underlying values of $\nu$, $\aN$, $\bN$, etc., but rather the values which best estimate the EV PDF.
\begin{figure}
	\includegraphics[width=1.0\textwidth]{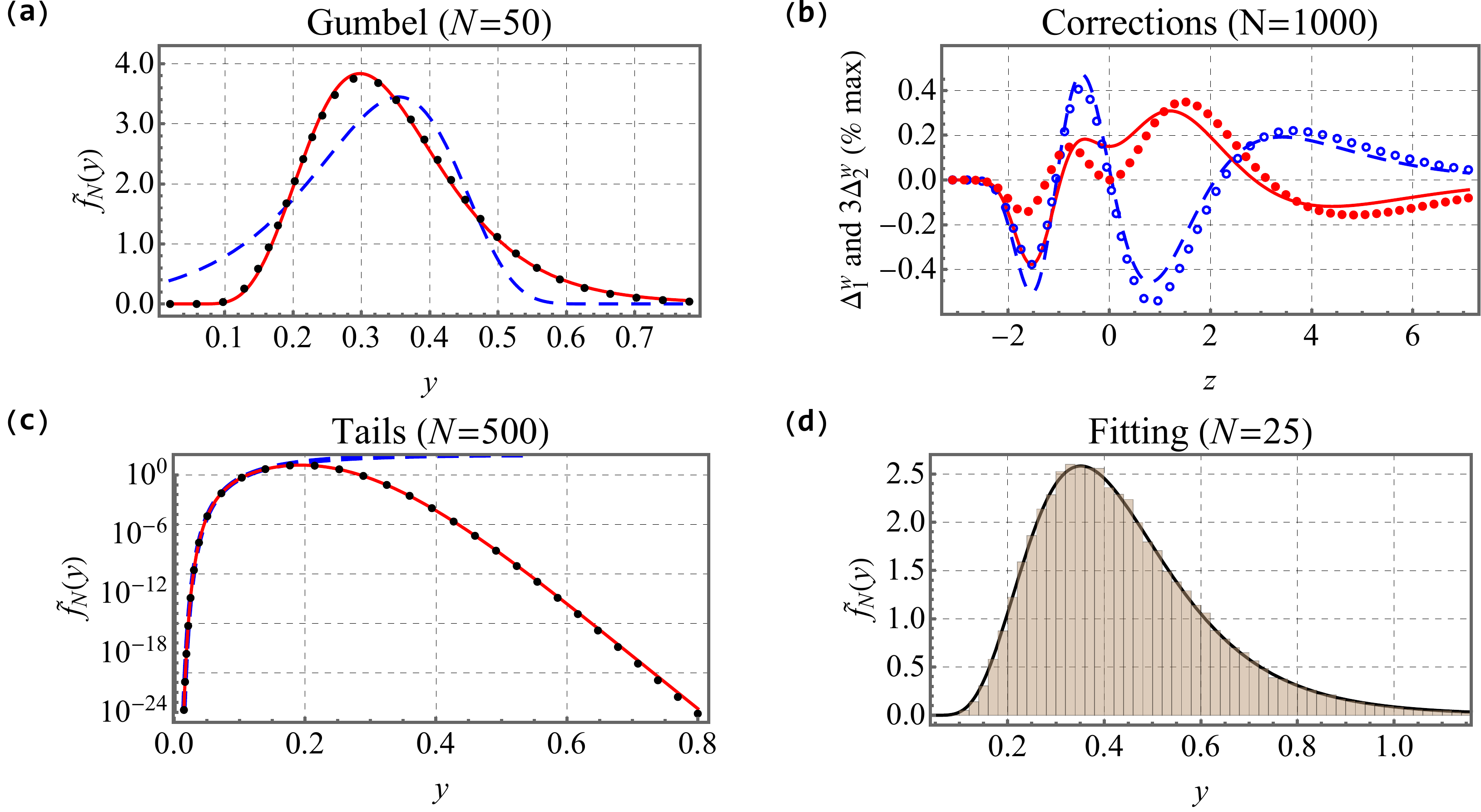}
	\caption{The case of the minimum value $y$ of $N$ bounded IID random variables $\chi\in[0,\infty)$, illustrated via an example of the one-dimensional first-return time problem, for which $\tilde{F}(\chi) = \text{erf}(1/\sqrt{\chi})$. Each panel corresponds to one of the previously discussed figures, and legends analogy is implied. (a) The zero-order $g_{\infty}(z)$ with the Lambert scaling and variables change method holds well to the exact values for $N=50$. Also shown is the approximation taken from \cite{Lawley}. See Fig.~\ref{figure: minimum case larger} in appendix \ref{section: larger} for a replot of (a) using a larger $N$. (b) The first correction to the transformed Gumbel approximation $\Delta_1^w$ has a magnitude of $\approx 0.5\%$ with respect to the maximal value of $g_{\infty}(z)$, $1/e$, and follows well its predicted shape. The second correction $\Delta_2^w$ is multiplied by $3$ for visibility, and follows its predicted shape. We used the exact values of $\aN^w$, $\bN^w$, $c_2^w$, and $c_3^w$. See Fig.~\ref{figure: minimum case larger} in appendix \ref{section: larger} for the non-transformed Gumbel corrections, namely $\Delta_1$ and $\Delta_2$, which are of a much larger magnitude. (c) The left tail and the uniform approximation for the PDF of $y$ with $N=500$ both excellently match the exact values. The uniform approximation functions for all $y$. Note that the left tail of a minimum EV problem is analog to the right tail of the maximum case. These problems have a trivial left large deviation function, and so it is omitted from the plot. (d) The practical method with $10^5$ minima of $N=25$ IID random variables each. Fit parameters of the zero-order were obtained from the data set and used to extract estimates for $g_{\infty}(z)$ which excellently match the samplings without assumed knowledge of the underlying distribution.}
\label{figure: minimum case}
\end{figure}

\section{Connection to other cases}
\label{section: others}

\subsection{Exceptional bounded distributions}

Usually, compact distributions lie in the Weibull universality class. However, when the PDF vanishes faster than a power-law at the endpoint, the asymptotic distribution is still Gumbel. An example of this is found in a problem discussed by Lawley~\cite{Lawley}, namely the minimum first-passage time to the origin of $N$ particles diffusing on the interval $(0,1)$ which start at the right reflective boundary, where the diffusion coefficient is $1/4$. Indeed, Lawley showed that in this case the Lambert W-function can be used to approximate $\aN$ and $\bN$, however, this case is included among those discussed above where $L$ is far from linear around $\aN$ for reasonably large $N$, and thus just using the Lambert representation of $\aN$ and $\bN$ is insufficient, and the change of variables must be employed as well. Since here we are dealing with a minimum rather than a maximum, the role of the CDF $F$ is replaced by the complementary CDF, $\tilde{F}(\chi) \equiv 1-F(\chi)$, given by $\tilde{F}(\chi) = \text{erf}(1/\sqrt{\chi})$ for this case, where $\chi\in[0,\infty)$ and $\text{erf}(\cdot)$ is the error function. Note that the exact complementary CDF for the minimum $y \equiv \min(\{\chi_1,...,\chi_N\})$ of any $N$ IID random variables is $\tilde{F}_N(y) = \tilde{F}^N(y)$. In what follows, we denote quantities of the minimum EV case by tildes.

As elaborated above, one needs to consider a variable change that renders the underlying CDF exponential-like. Observing the asymptotic behavior $\tilde{F}(\chi) \simeq 1-\sqrt{\chi/\pi}\exp(-1/\chi)$, it is clear that the relation must be $\omega = 1/\chi$, then for large $\omega$ one has $L_{\omega}(\omega) \propto \omega + O[\ln(\omega)]$, which is very close to being linear and therefore can be Taylor approximated very well. Since the minimum $y$ of $\{\chi_i\}$ is the maximum $w$ of $\{\omega_i\}$, we can apply our procedures of the above sections to the maximum value problem whose CDF is $F_{\omega}(\omega) = \tilde{F}(1/\omega) = \text{erf}(\sqrt{\omega})$. The PDF for $y$, $\tilde{f}_N(y)$, is then obtained from the PDF for $w$, $f_N^w(w)$, by
\begin{equation}
\label{equation: general first two corrections change variables}
    \tilde{f}_N(y) = \frac{1}{y^2}f_N^w\left(\frac{1}{y}\right) .
\end{equation}
The results of our above discussions for this example are presented in Fig.~\ref{figure: minimum case}, with each panel demonstrating a previous figure: (a) Fig.~\ref{figure: gumbel zero}, (b) Fig.~\ref{figure: corrections change}, (c) Fig.~\ref{figure: tails approx}, and (d) Fig.~\ref{figure: gumbel fit}. The agreement is indeed excellent, and the key is that $f_N^w(w)$ is much closer to a Gumbel distribution than $\tilde{f}_N(y)$, similarly to the super-compressed and super-stretched cases. In appendix \ref{section: larger}, we replot panel (a) with an $N$ that is larger by a factor of $10^2$, see Fig.~\ref{figure: minimum case larger}, demonstrating the slow rate of convergence. We also show the corrections to the non-transformed Gumbel case, in analogy to Fig.~\ref{figure: corrections delta}, where a magnitude of $\approx 20\%$ can be seen (in contradiction to $0.5\%$ for the transformed case). As far as data analysis is concerned, one simply needs to employ the algorithm seen in Fig.~\ref{figure: gumbel chart} for $\nu<0$, while sorting descendingly instead of ascendingly in step (i). As for the moments, we have
\begin{equation}
\label{equation: moments definition minimum}
	\left<y^m\right> \equiv \int_0^{\infty}\text{d}y\,\tilde{f}_N(y)y^m = \int_0^{\infty}\text{d}w\,f_N^w(w)w^{-m} ,
\end{equation}
so the $m$th moment of $y$ is just the $-m$ of $w$, which we have already calculated above. Our right tail and uniform approximations for the distribution of $w$ immediately yields the left tail and uniform approximations of $y$'s distribution.

\subsection{Other extreme value limits}

As a final remark, we point out a nice observation for the reason why random variables with EV distribution different than Gumbel do not suffer from the poor logarithmic convergence problems of their Gumbel counterparts. Take, for example, a distribution with a power-law tail, $f(\chi) \propto \chi^{-\mu}$ with $\mu>0$. A direct application of the method used here would have us expand $L(\chi) \simeq \mu\ln(\chi)$ around $\aN \propto N^{1/\mu}$, obtaining $\bN \simeq \aN/\mu$. As a consequence, $c_n \sim O(1)$ for any $n>1$, and so all terms in the expansion of $L$ are of the same order, resulting in the Gumbel universality being lost. The same is true for a compact distribution with $f(\chi) \propto (1-\chi)^{\mu-1}$.
\begin{table}
	\begin{tabular}{|c|c|c|c|c|c|c|}
		\hline
		Type & Support & CDF & Extreme value & Scaling sequences & Limit & Limiting CDF \\ \hline
		Power-law & $\chi_{\rm p}\in[1,\infty)$ & $1 - \chi_{\rm p}^{-\mu}$ & $x_{\rm p}=\aN^{\rm p}+\bN^{\rm p}z_{\rm p}$ & $\aN^{\rm p}=N^{1/\mu}$ , $\bN^{\rm p}=\mu^{-1}N^{1/\mu}$ & Fr\'echet & $\exp[-(1+z_{\rm p}/\mu)^{-\mu}]$ \\ \hline
		Compact & $\chi_{\rm c}\in[0,1]$ & $1-(1-\chi_{\rm c})^{\mu}$ & $x_{\rm c}=\aN^{\rm c}+\bN^{\rm c}z_{\rm c}$ & $\aN^{\rm c}=1-N^{-1/\mu}$ , $\bN^{\rm c}=\mu^{-1}N^{-1/\mu}$ & Weibull & $\exp[-(1-z_{\rm c}/\mu)^{\mu}]$ \\ \hline
		Exponential & $\chi_{\rm e}\in[0,\infty)$ & $1 - \exp\left(-\mu\chi_{\rm e}\right)$ & $x_{\rm e}=\aN^{\rm e}+\bN^{\rm e}z_{\rm e}$ & $\aN^{\rm e}=\mu^{-1}\ln(N)$ , $\bN^{\rm e}=\mu^{-1}$ & Gumbel & $\exp(-e^{-z_{\rm e}})$ \\ \hline
		Gaussian & $\chi_{\rm g}\in(-\infty,\infty)$ & $[1+\text{erf}(\chi_{\rm g}/\sqrt{2})]/2$ & $x_{\rm g}=\aN^{\rm g}+\bN^{\rm g} z_{\rm g}$ & $\aN^{\rm g},\bN^{\rm g}$ & Gumbel & $\exp(-e^{-z_{\rm g}})$ \\ \hline
	\end{tabular}
	\caption{The considered random variables, where $\mu>0$ is a constant parameter.}
\label{table: random variables classes}
\end{table}

It is instructive to look at this from the perspective of a change of variables. Let us consider the four random variables that appear in table~\ref{table: random variables classes}. The transformations
\begin{equation}
\label{equation: variables transformations to exp 1}
	\chi_{\rm e} = \ln\left(\chi_{\rm p}\right) , \quad \chi_{\rm e} = \ln\left( \frac{1}{1-\chi_{\rm c}} \right) , \quad \chi_{\rm e} = - \frac{1}{\mu} \ln\left[ \frac{1}{2}-\frac{1}{2} \text{erf}\left(\frac{\chi_{\rm g}}{\sqrt{2}}\right) \right] ,
\end{equation}
generate the exponentially distributed random variable from the power-law, compact, and Gaussian variables, respectively. Since these are strictly increasing functions, Eq.~(\ref{equation: variables transformations to exp 1}) holds for the EVs as well. When plugging these in, Eq.~(\ref{equation: variables transformations to exp 1}) yields
\begin{equation}
\label{equation: variables transformations to exp 2}
	z_{\rm e} = \mu\ln\left(1+\frac{z_{\rm p}}{\mu}\right) , \quad z_{\rm e} = -\mu\ln\left(1-\frac{z_{\rm c}}{\mu}\right) , \quad z_{\rm e} = z_{\rm g} - \ln(N) + \frac{1}{2} \left(\aN^{\rm g}\right)^2 + \ln\left( \sqrt{2\pi} \aN^{\rm g} \right) + O\left[\frac{1}{\left(\aN^{\rm g}\right)^2}\right] ,
\end{equation}
with $\bN^{\rm g}=1/\aN^{\rm g}$ and $\aN^{\rm g} \gg 1$ for the Gaussian case. Note that for the first two cases the $N$ dependency vanishes from the relation between the rescaled variables. Moreover, plugging $z_{\rm e}$ in terms of $z_{\rm p},z_{\rm c}$ into the Gumbel CDF results in the Fr\'echet and Weibull CDFs, respectively. Thus, the power-law and compact random variables are actually an exponentially distributed variable in another guise. Hence, it is not surprising that the convergence rate to these limits is much faster, as for the exponential case all finite-$N$ corrections to the Gumbel limit vanish. The latter statement can be concluded by plugging $L(\chi) \simeq \mu\chi$ into Eq.~(\ref{equation: normalization sequences definition}), which yields $c_n=0$ for $n>1$. However, for the Gaussian case in Eq.~(\ref{equation: variables transformations to exp 2}) things are different, as the $N$ dependency remains. Actually, if we identify $z_{\rm e}=z_{\rm g}$, Eq.~(\ref{equation: variables transformations to exp 2})'s rightmost section exactly reproduces Eq.~(\ref{equation: normalization sequences zero order}), with appropriate Gaussian parameters. This further emphasizes the naturalness of the Lambert scaling approach for distributions yielding the Gumbel limit when $N\to\infty$.

\section{Summary}
\label{section: summary}

In this paper, we have discussed the EV problem of $N$ IID random variables and constructed a theory that makes the Gumbel limit of the EV distribution usable for values of $N$s below $500$, and in most cases less than a hundred, whereas in some cases the standard approach would completely fail for $N$s which are not astronomically large. Exploiting the Lambert W-function, we obtained the scaling sequences $\aN$ and $\bN$ as simple asymptotic series in terms of a single parameter $\alphaN$, see Eq.~(\ref{equation: normalization sequences zero order}). The expansions obtained generate useful approximations (sometimes with the aid of Pad\'e transformation) down to $N=50$. Applying a simple variable transformation makes the Gumbel limit relevant in its uncorrected form, namely $g_{\infty}(z)$. We also provided a simple way to derive arbitrary-order corrections to the Gumbel distribution for the EV of IID random variables, and demonstrated the first two corrections. We have tested this for a whole family of stretched or compressed exponential distributions, including the slowly-converging super-stretched case. We improved the accuracy of the large-deviation representation of the right tail of the EV distribution while allowing for a uniform approximation that captures the close left tail as well. If the underlying distribution is not given, we described a fitting scheme that yields an excellent match between a given data set and the Gumbel limit. We have also shown how the same techniques works for compact distributions with essential singularities at the endpoint of the distribution.

\begin{acknowledgments}

The support of the Israel Science Foundation, Grant No. 1898/17, is acknowledged.

\end{acknowledgments}

\appendix
\setcounter{figure}{0}
\renewcommand{\thefigure}{A\arabic{figure}}

\section{Supporting figures}
\label{section: larger}

This appendix contains an analog to Fig.~\ref{figure: gumbel zero}, replotted with $N$ that is $10^3$ times larger than the one used for its main text counterpart, to demonstrate how slow the convergence rate really is. The same is done for panel (a) of Fig.~\ref{figure: minimum case}, this time with a factor of $10^2$. As the existing theory already uses the Lambert W-function in this case, here the increase in convergence rate due to our theory originates mainly from the change of variables method. We also add an analog of Fig.~\ref{figure: corrections delta} for the minimum EV, which shows that also for this case the magnitude of the corrections prior to transforming variables is much larger ($20\%$ compared to $0.5\%$ after making the change of variables).
\begin{figure}[ht]
	\includegraphics[width=1.0\textwidth]{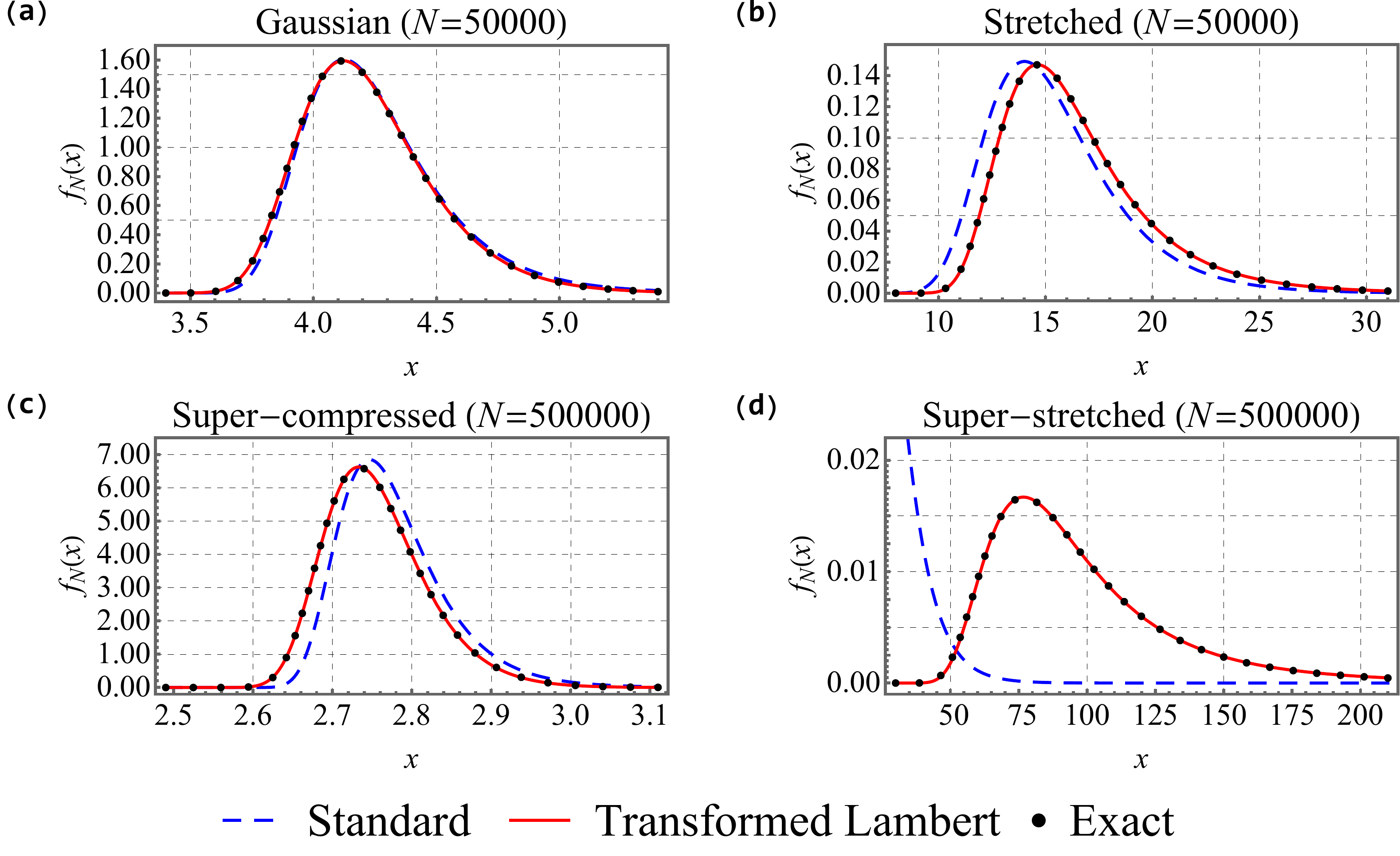}
	\caption{An analog to Fig.~\ref{figure: gumbel zero}, with an $N$ that is larger by a factor of $10^3$. Except for the Gaussian, the standard approximation of all cases converges very slowly.}
\label{figure: gumbel zero larger}
\end{figure}
\begin{figure}[ht]
	\includegraphics[width=1.0\textwidth]{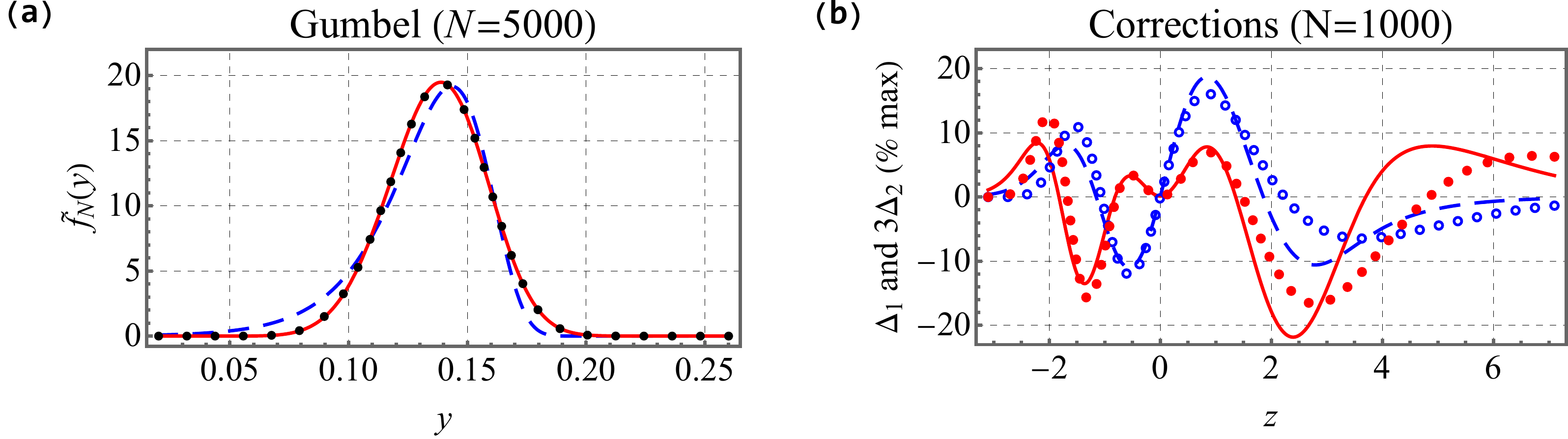}
	\caption{(a) An analog to panel (a) of Fig.~\ref{figure: minimum case}, with an $N$ that is larger by a factor of $10^2$. (b) An analog to Fig.~\ref{figure: corrections delta} for the minimum case. The magnitude of the corrections is much larger before making the transformation $w=1/y$.}
\label{figure: minimum case larger}
\end{figure}

\end{document}